\documentclass[aps,prd,english,superscriptaddress,11pt,notitlepage]{revtex4}
	\usepackage[colorlinks=true, a4paper=true, pdfstartview=FitV,
linkcolor=blue, citecolor=blue, urlcolor=blue]{hyperref}

\usepackage{amsmath}
\usepackage{amssymb}
\usepackage{graphicx}
\usepackage{hhline}
\usepackage{textcomp}
\makeatletter
\usepackage{babel}
\newcommand{\bea}{\begin{eqnarray}}
\newcommand{\eea}{\end{eqnarray}}

\newcommand{\be}{\begin{equation}}
\newcommand{\ee}{\end{equation}}
\usepackage[active]{srcltx}
\begin{document}


 \title{Background fields and self-dual Skyrmions}
 
 \author{C. Naya}
\affiliation{INFN, Sezione di Lecce, Via per Arnesano, C.P. 193 I-73100 Lecce, Italy}

\author{K. Oles}
\affiliation{Institute of Physics,  Jagiellonian University, Lojasiewicza 11, Krak\'{o}w, Poland}
 
\begin{abstract}
We show that a suitable background field can bring a non-BPS topological soliton into its BPS, {\it self-dual}, counterpart. As an example we consider Skyrmions in the minimal Skyrme model. We prove the triviality of the corresponding moduli space. This means that the resulting self-dual Skyrmion does statically interact with the background field. 

We also show that the originally self-dual Skyrmions (e.g. solutions of the BPS Skyrme model) can preserve the self-duality after a coupling with a background field. In this case, BPS Skyrmions can be effortless moved with respect to the background. 

\end{abstract}

\maketitle

\section{Motivation}

A detailed understanding of interactions of topological solitons, that is, localized particle-like solutions of nonlinear field equations, which carry a nontrivial value of a topological charge \cite{MS,S}, is a great challenge of the contemporary physics. It is important not only from theoretical reasons providing a deeper insight into dynamical properties of solitons at classical as well as quantum level, which ultimately may allow for an explanation of various phenomena occurring during scattering, annihilation and creation processes, but also due to possible practical applications in a number of condense matter materials supporting them. Indeed, a manipulation of topological solitons, their creation (annihilation) as well as dynamical stabilisation, is crucial for any realistic application. 

Unfortunately, except in integrable theories, we have a rather limited understanding of dynamics of topological solitons even at the qualitative stage. A general method, known as the moduli space approximation, works quite well if, in the leading order, a given solitonic process occurs along a geodesic flow. It means that the process happens via a sequence of energetically equivalent, self-dual (SD) states, where transition between the states is triggered by a corresponding zero mode. In the next step one may take into account also massive (bound) modes which, by coupling to the zero mode, introduce the so-called Coriolis forces. The obtained effective model (where we are left with a finite number of degrees of freedom corresponding basically to amplitudes of the modes) quite well explains various aspects of the dynamics of the process in question. 
This construction was further extended to processes which do not support an infinite number of energetically equivalent states. Then, the moduli space has to be replaced by the so-called unstable manifold \cite{unstable}, which could be treated as a moduli space with a drag force (effective potential).  Unfortunately, while this extension applies to soliton-soliton collisions, it completely fails for soliton-antisoliton (SAS) scatterings. In consequence, even the simplest kink-antikink collision in the $\phi^4$ model in (1+1) dimensions lacks an explanation \cite{phi4}. 

However, recently a method which transforms an SAS solution of a given model $L[\phi]$ into a self-dual counterpart has been proposed. It is based on an addition of a non-dynamical background field $\sigma$ (impurity) which couples to the original theory in a particular self-dual manner \cite{SD-imp-1}-\cite{SD-imp-3}
\be
L [\phi] \rightarrow L[\phi, \sigma].
\ee
Effectively, the introduction of the self-dual background field switches off (screens) the static inter-soliton forces and brings the considered SAS state into a self-dual configuration. This means that it solves a corresponding Bogomolny equation and a moduli space exists.  For example, this construction provides self-dual kink-antikink solutions in the $\phi^4$ model in (1+1) dimension \cite{SD-imp-3}. In consequence, for such a self-dual deformed model, the lowest order annihilation (scattering) process occurs as a geodesic flow on a certain moduli space. This allowed for a systematical understanding of the role of internal modes \cite{SD-imp-4} (which in the deformed model nontrivially depend on the position on the moduli space) in SAS  dynamics leading to the discovery of spectral walls \cite{SW}, \cite{thick-SW}. 

The importance of this self-dual background field framework is related to the fact that it can be applied to any multi-solitonic scattering provided  the initial as well as the final states are {\it self-dual solitons}. It is always the case for one scalar field theory in (1+1) dimensions, where a static (anti)soliton is a solution of a first order Bogomolny equation. However, most of topological solitons in higher dimensions do not enjoy such a self-dual property. Hence, we cannot use the self-dual background field framework as a tool for the analytical understanding of SAS dynamics.

This is the aim of the present work to show that a non-self-dual (non-BPS) soliton can be made a self-dual one by means of another background field $\Omega$, which transforms a possible asymptotic state into a SD solution
\be
 L [\phi] \rightarrow L[\phi, \Omega].
\ee
It is worth mentioning that another well-known option to obtain BPS configurations is by changing the geometry of the base space. For instance, this has been applied to Skyrmions in $S^3$ \cite{Manton1986} or vortices in the hyperbolic plane $\mathbb{H}^2$ \cite{Witten1977,Paul2012} (see \cite{MS}, Chapter 7, for a detailed discussion of the latter).

Hence, this work is the first step towards our ultimate goal, which is to enlarge the applicability of the self-dual background field framework to solitonic collisions between asymptotically non-BPS states, i.e., basically to any process in any field theory which schematically can be described as
\be
 L [\phi] \rightarrow L[\phi, \Omega] \rightarrow L[\phi, \Omega, \sigma].
\ee
Here the first background field transforms a non-SD asymptotical state into a SD soliton, while the second background field puts an SAS pair on a moduli space.
As a particular example we consider the minimal Skyrme model in (3+1) dimensions. 

\section{The minimal Skyrme model}
The static energy functional of the minimal Skyrme model \cite{Skyrme} consists of two terms only (in dimensionless units, with the energy and length units rescaled) 
\be
E=E_2+E_4,
\ee
where $E_2$ is the Dirichlet energy,
\be
E_2=\int_{\mathbb{R}^3} -\frac{1}{2} \mbox{ Tr } (R_i R_i) d^3 x,
\ee
while $E_4$ is the Skyrme term,
\be
E_4=\int_{\mathbb{R}^3} -\frac{1}{16} \mbox{ Tr } ([R_i, R_j] [R_i,R_j]) d^3 x,
\ee
needed to circumvent Derrick's theorem so stable solutions may exist.

Here $R_i = \partial_i U U^{-1}$ is the right invariant current and $U(\vec{x})$ is an $SU(2)$ valued matrix field.  It is a very well known result that the model has a topological bound derived by Faddeev \cite{F}. It can be shown by using the eigenvalues $\lambda_i^2$ of the strain tensor, which is the three dimensional symmetric positive matrix $
D_{ij}=-\frac{1}{2} \mbox{ Tr } (R_i R_j)
$ \cite{NM-lambda}. 
Then, the energy can be rewritten as
\bea
E&=&\int_{\mathbb{R}^3} ( \lambda^2_1+\lambda^2_2+\lambda^2_3 + \lambda^2_1\lambda^2_2+\lambda^2_2\lambda^2_3 + \lambda^2_3\lambda^2_1)d^3 x \nonumber \\
&=& \int_{\mathbb{R}^3} \left(  ( \lambda_1 \pm  \lambda_2  \lambda_3)^2 + ( \lambda_2 \pm  \lambda_3  \lambda_1)^2 + ( \lambda_3 \pm  \lambda_1 \lambda_2)^2 \right) d^3 x \mp 6 \int_{\mathbb{R}^3} \lambda_1 \lambda_2 \lambda_3 d^3 x  \nonumber \\
& \geq & 6 \left| \int_{\mathbb{R}^3} \lambda_1 \lambda_2 \lambda_3 d^3 x \right| = 12 \pi^2 |B|,
\eea
where we have used that the density $\mathcal{B}_0$ of the topological charge $B$ reads
\be
\mathcal{B}_0 = \frac{1}{2\pi^2} \lambda_1 \lambda_2 \lambda_3 \;\;\; \Rightarrow \;\;\; B= \int_{\mathbb{R}^3} \mathcal{B}_0  d^3 x .
\ee
However, the bound cannot be saturated on the $\mathbb{R}^3$ base space. Indeed, the Bogomolny equations 
\be
\lambda_1 =  \pm \lambda_2  \lambda_3, \;\; \lambda_2 = \pm  \lambda_3  \lambda_1, \;\; \lambda_3 =  \pm \lambda_1  \lambda_2, \label{BOG-Sk}
\ee
imply that $\lambda_1^2=\lambda_2^2=\lambda_3^2=1$, which has no topologically nontrivial solution on $\mathbb{R}^3$ \cite{NM-lambda}. In  consequence, minimal Skyrmions are not BPS configurations. In fact, they are rather strongly bounded solitons, which was a source for one of the main problems as the application of the model to atomic nuclei is concerned. Namely, the appearance of nonphysically large binding energies. To circumvent this issue it is necessary to departure from the minimal Skyrme model and add new but physically very well motivated terms. One option is to remain with the same field content and add the sextic (topological current squared) term \cite{BPS-Sk,PRL} or a so-called lightly bound potential \cite{DH,Sp, Gud-1} (or mixture of both \cite{Gud}). Another possibility is to couple (infinitely many) vector mesons \cite{vec-Sk-1}-\cite{vec-Sk-3}.

\section{Background field and Skyrmions }
\subsection{Topological bound, Bogomolny equations and moduli space}
The non-existence of nontrivial solutions of the Bogomolny equations (\ref{BOG-Sk}) means that this system of first order differential equations is {\it too restrictive}, but it can be relaxed if a suitable background field is coupled. In this context, we would like to remark that the first background field deformation
of the minimal Skyrme model leading to BPS Skyrmions was presented in \cite{LF} and \cite{LF-YS}. We further comment on this model later, when some similarities are discussed.

For the sake of generality, we start with a triplet of background functions (impurities) $\Omega_1 (\vec{x}), \Omega_2(\vec{x}), \Omega_3(\vec{x})$. Then, a background field deformation of the minimal Skyrme model we are going to focus on has the form
\be
E_\Omega = E_{2, \Omega} + E_{3, \Omega}  + E_4 \label{model},
\ee
where $E_{2, \Omega} $ is the deformed Dirichlet energy,
\be
E_{2, \Omega} = \int_{\mathbb{R}^3} \left( (1+\Omega_1)^2 \lambda^2_1+(1+\Omega_2)^2 \lambda^2_2+(1+\Omega_2)^2 \lambda^2_3 \right)d^3 x,
\ee
and $E_{3, \Omega}$ is the topological charge integral in the presence of the background fields,
\be
E_{3, \Omega} = -2 \int_{\mathbb{R}^3}  (\Omega_1+\Omega_2+\Omega_3) \lambda_1 \lambda_2 \lambda_3 d^3 x,
\ee
which for non-constant background fields is not a purely topological (boundary) term. The Skyrme term remains unchanged. Of course, the addition of the background fields breaks the translational symmetry explicitly.

The motivation for this particular form of the background deformed solitonic models comes from former work on the self-dual background fields in (1+1) dimensions \cite{SD-imp-1}, \cite{SD-imp-2}. In fact, such a deformation required a modification of one of the original terms in the Lagrangian, the standard kinetic term (two derivatives) or the potential (no derivatives). For simplicity, the lower derivative term, i.e., the potential, was typically multiplied by a background field. Here, we follow this pattern and also deform the term with the lower number of derivatives, that is, the Dirichlet energy. In addition, it was necessary to include a background field deformation of the topological term. Here it is simple represented by a multiplication of the baryon charge density by a background field function.

Now, the topological bound coincides with the Faddeev bound
\bea
E_\Omega &= &\int_{\mathbb{R}^3} \left( ( (1+\Omega_1)\lambda_1 -  \lambda_2  \lambda_3)^2 + ( (1+\Omega_2)\lambda_2 -  \lambda_3  \lambda_1)^2 + ( (1+\Omega_3)\lambda_3 -  \lambda_1 \lambda_2)^2  \right) d^3 x \nonumber \\
&+& 6 \int_{\mathbb{R}^3} \lambda_1 \lambda_2 \lambda_3 d^3 x \nonumber \\
& \geq &  6  \int_{\mathbb{R}^3} \lambda_1 \lambda_2 \lambda_3 d^3 x  = 12 \pi^2 B.
\eea
However, the corresponding Bogomolny equations differ and, at least for some background fields, may allow for Skyrmions
\be
 (1+\Omega_1)\lambda_1 =  \lambda_2  \lambda_3, \;\;  (1+\Omega_2) \lambda_2 =   \lambda_3  \lambda_1, \;\;  (1+\Omega_3)\lambda_3 =  \lambda_1  \lambda_2. \label{BOG-Sk-Imp}
\ee
Indeed, after simple manipulations we find that 
\be
\lambda^2_1 = (1+\Omega_2)(1+\Omega_3), \;\;\; \lambda^2_2 = (1+\Omega_3)(1+\Omega_1), \;\;\;  \lambda^2_3 = (1+\Omega_1)(1+\Omega_2),
\ee
which, depending on a particular form of the background fields, may have a topologically nontrivial solution on $\mathbb{R}^3$.  

Note that for any solution of the Bogomolny equations the topological bound is saturated and the energy density $\mathcal{E}$ and baryon charge density $\mathcal{B}_0$ coincide (up to an irrelevant numerical factor)
\be
\mathcal{E} = 12 \pi^2 \mathcal{B}_0= 6 \lambda_1\lambda_2\lambda_3.
\ee
Using the Bogomolny equations and choosing the plus sign, we find that 
\be
\mathcal{E} = 12 \pi^2 \mathcal{B}_0=  6 (1+\Omega_1)(1+\Omega_2)
(1+\Omega_3).
\ee
Surprisingly, the physical densities in the self-dual sector are {\it fixed} by the background fields. Thus, the corresponding moduli space is trivial, i.e., a BPS soliton, if it exists, can be located only at a certain spatial point. In other words, the distance between the BPS Skyrmion and the impurity is fixed. This means that the Skyrmion and the impurity do statically interact, even though the resulting bound state saturates the topological bound. Hence, the translational symmetry is not restored in the self-dual sector. All that should be contrasted with the former findings in self-dual background field deformations, where soliton and impurity did not statically interact \cite{SD-imp-1}-\cite{SD-imp-3}. This had given rise to a nontrivial moduli space, physically describing the soliton and impurity at any distance from each other. We will further investigate this issue below. 
\subsection{Isotropic background fields and charge one BPS Skyrmion}

As the simplest possibility we consider the {\it isotropic} case when all background field functions are the same $\Omega_1=\Omega_2=\Omega_3\equiv \Omega$. This implies that also the eigenvalues coincide, $\lambda^2_1=\lambda^2_2=\lambda^2_3 \equiv \lambda^2$. Hence, the Bogomolny equations (\ref{BOG-Sk-Imp}) are reduced to one condition
\be
\lambda^2 = (1+\Omega)^2 \label{BOG-uniform}.
\ee
To verify the existence of a topologically nontrivial solution of the deformed Bogomolny equation, we apply the usual parametrisation of the Skyrme field 
\be
U=\exp (i \xi (\vec{x}) \vec{\tau} \cdot \vec{n}(\vec{x})), \label{U}
\ee
where $\vec{\tau}$ are the Pauli matrices, $\xi$ is a real valued function, while $\vec{n}$ is a unit three-component vector, typically expressed via the stereographic projection by a complex field $u \in \mathbb{C}$ 
\be
\vec n = \frac{1}{1+|u|^2} \left(2 \Re(u), 2\Im(u),1-|u|^2\right). \label{n}
\ee
Furthermore, we use spherical coordinates $(r,\theta, \phi)$ and assume that $\xi=\xi(r)$ and $u=u(\theta, \phi)$, which includes e.g. the usual hedgehog solution for the charge one Skyrmion. Then, the eigenvalues can be found as  
\be
\lambda_1^2=\xi_r^2, \;\;\;
\lambda_2^2+\lambda_3^2=4\sin^2 \xi \frac{\nabla u \cdot \nabla \bar{u}}{(1+|u|^2)^2} .
\ee
The assumed equality of the eigenvalues gives
\be
\xi_r^2=2\sin^2 \xi \frac{\nabla u \cdot \nabla \bar{u}}{(1+|u|^2)^2} .
\ee
This equation has the following solution with the unit topological charge
\be
\xi=2\arctan \frac{r_0}{r}, \;\;\; u=e^{i\phi} \tan \frac{\theta}{2}, \label{sol-uniform}
\ee
where $r_0>0$. Now, the full system is well defined only if the background field is related to the eigenvalue by the Bogomolny equation (\ref{BOG-uniform}). This requires that the background field must be of the form
\be
1+\Omega (r)= \frac{2r_0}{r_0^2+r^2}. \label{Omega-1}
\ee
This is the unique (up to the positive parameter $r_0$) impurity of  the isotropic type which provides the self-dual hedgehog-type Skyrmion. In other words, for this background field the deformed minimal Skyrme model is a BPS theory with the $B=1$ Skyrmion (of the hedgehog geometry) saturating the topological bound. For another spatial dependence of the isotropic background field $\Omega$ our model does not support self-dual solitons carrying unit topological charge. 

We remark that the condition that all eigenvalues are the same holds also for the solutions of the Bogomolny equations coming from a version of the Skyrme model proposed by D. Harland \cite{DH}. Hence, the background field plays the role of the potential present in \cite{DH} in the resulting Bogomolny equations. Of course, both models have exactly the same $B=1$ solution, although in our set-up the scaling parameter $r_0$ is fixed by a particular form of the impurity. Note that the bound in the background field deformed model (\ref{Omega-1}) can be attained only for $B=0$ and $B=1$ solutions. 

On the other hand, the isotropic Bogomoly equations (\ref{BOG-uniform}) are identical to the Bogomolny equations previously obtained in the background field deformed Skyrme model considered in \cite{LF}, \cite{LF-YS}. Therefore, the solution for the isotropic background field (\ref{sol-uniform}) must reproduce the solution found in \cite{LF}. Apparently, in spite of different energy functionals, both background deformations have exactly the same self-dual sector sharing the BPS solutions.

It is also possible to find out background fields in which the $B=1$ solution of the minimal Skyrme model becomes self-dual. For that we need to use the following set of non-isotropic background fields
\be
1+\Omega_2=1+\Omega_3\equiv 1+\Omega(r), \;\;\; 1+\Omega_1=(1+\Omega(r))\mu(r),
\ee
where both $\Omega$ and $\mu$ are functions of the radial variable $r$ which remain to be determined. Then, for the hedgehog solution we get
\be
\xi_r^2=(1+\Omega)^2, \;\;\; \frac{\sin^2 \xi}{r^2} =(1+\Omega)^2\mu.
\ee
Hence, combining them together we arrive at $\sin^2 \xi = r^2 \xi^2_r \mu$. Assuming that the profile $\xi (r)$ is exactly the profile of the $B=1$ Skyrmion of the minimal Skyrme model, we can obtain the correct spatial form of $\mu(r)$. Finally, using the first Bogomolny equation $\xi^2_r=(1 + \Omega)^2$ we get the $\Omega$ background as well. 

\begin{figure}
\hspace*{-1.0cm}
\includegraphics[width=.8\textwidth]{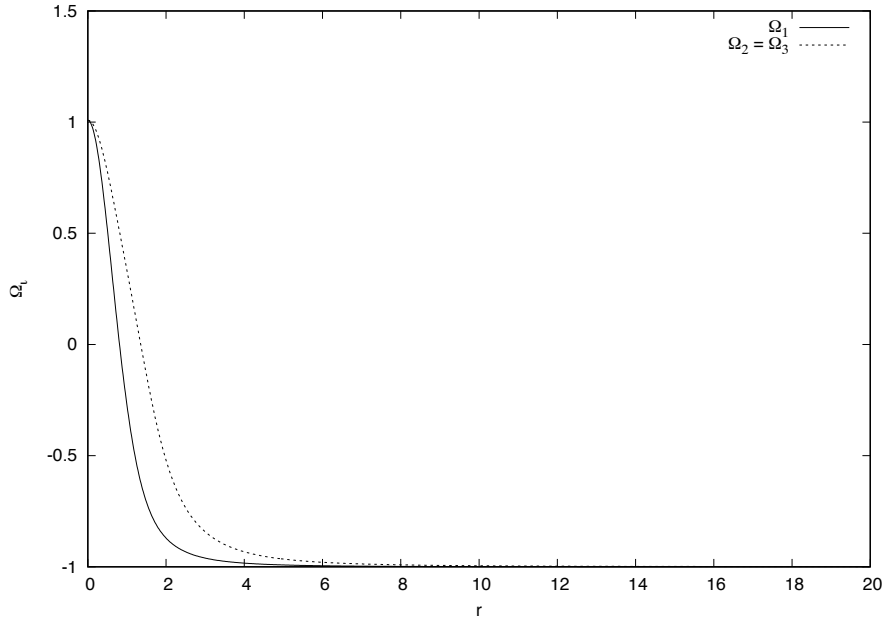}
\caption{The background fields which transform the $B=1$ soliton of the Skyrme model into a self-dual soliton.}
\label{hedgehog}
\end{figure}

The different background fields are presented in Fig. \ref{hedgehog}. For this purpose, a minimization of the energy functional without the impurity was performed in order to obtain the profile $\xi(r)$ of the hedgehog within the minimal Skyrme model and derive the background fields as explained. We used a numerical gradient flow method in a 1-dimensional lattice of 1000 points with an interspacing $\Delta r = 0.02$ and finite diference approximations of fourth order for the derivatives were implemented.

In our example the isotropic background field $\Omega$ must tend to -1 at spatial infinity. This means that all coefficients $(1+\Omega_i)^2$ which multiply terms in the Dirichlet energy tend to zero. One may ask the question whether it is possible to have background fields whose behavior at spatial infinity is more regular, i.e., $(1+\Omega_i) \nrightarrow 0$. However, we will present an argument that it is rather impossible, at least for spherically symmetric Skyrmions. With this purpose we assume that $(1+\Omega_1)\rightarrow a$ while $(1+\Omega_2)\rightarrow b$ and $(1+\Omega_3)\rightarrow b$, where $a,b \neq 0$. This means that, asymptotically, the Bogomolny equations are 
\be
\lambda^2_1=b^2, \;\;\; \lambda^2_2=\lambda^2_3=ab.
\ee
The spherical symmetry implies that $\xi_r=\pm b$. But this is in contradiction with the assumption that $\xi \rightarrow 0$ (or $\pi$) at infinity. 

\subsection{Triviality of the moduli space}

Typically, Bogomolny equations result in a whole family of energetically equivalent solutions which are parameterized by a set (finite or even infinite) of moduli space coordinates. It happens for usual Poincare invariant models supporting self-dual topological solitons: topological kinks in (1+1) dimensions (for example $\phi^4$ or sine-Gordon theories), lumps in the $O(3)$ $\sigma$-models in (2+1) dimensions, the Abelian Higgs model at the critical coupling and many others \cite{MS}. Furthermore, also after coupling with the self-dual background field (which breaks the translational symmetry at the level of the Lagrangian) this feature remains unchanged \cite{SD-imp-1}-\cite{SW} (see also \cite{Tong}-\cite{Krush-2} in the context of the Abelian Higgs model). Therefore, also here one could think that the equality of the eigenvalues gives a one-parameter family of the solutions (\ref{sol-uniform}). However, the background field chooses {\it only one} particular value of the parameter $r_0$. In a sense, the restoration of the scaling symmetry implied by the equality of $\lambda$s is again broken by the fixed form of the impurity. Equivalently, it shows that our solutions are the unique solutions with spherical symmetry and therefore represent {\it a Skyrmion on top of the impurity}. Nonetheless, having in mind (1+1) dimensional self-dual background field models, one could hope that a moduli space still may exist. In fact, in (1+1) dimensions the self-dual background field deformed models always have a moduli space of solutions related to the (generalized) translation invariance \cite{SD-imp-2}-\cite{SD-imp-3}. For example, it can describe a kink and antikink at any distance from each other \cite{SD-imp-3}. However, as we have already noticed it is not the case for the current construction. 

To better analyse this issue, let us consider the simplest isotropic case. As all background functions are the same, we arrive at a Skyrme-type model where the Dirichlet term as well as the topological term are simply multiplied by functions of $\vec{r}$. Hence, we can apply the standard parametrization and find 
\bea
E&=&\int_{\mathbb{R}^3} d^3x \; (1+\Omega)^2 \left(  \xi_i^2 + 4\sin^2\xi \frac{u_i\bar{u}_i}{(1+|u|^2)^2} \right) \nonumber \\
&+& \int_{\mathbb{R}^3} d^3x \left(4\sin^2\xi \left( \frac{\xi_i^2 u_i\bar{u}_i}{(1+|u|^2)^2} -\frac{\xi_i u_i \xi_j\bar{u}_j}{(1+|u|^2)^2} \right)+ 4\sin^4 \xi \frac{(u_i\bar{u}_i)^2-u_i^2\bar{u}_j^2}{(1+|u|^2)^4} \right) \nonumber \\
&-& 2 \cdot 3\int_{\mathbb{R}^3} d^3x \; \Omega \frac{2i\sin^2 \xi}{(1+|u|^2)^2} \epsilon_{ijk}\xi_iu_j\bar{u}_k.
\eea
Then, it can be rewritten as
\bea
E&=&\int_{\mathbb{R}^3} d^3x \left( (1+\Omega) \xi_i -2i\frac{\sin^2 \xi}{(1+|u|^2)^2} \epsilon_{ijk}u_j\bar{u}_k\right)^2 \nonumber \\
&+&\int_{\mathbb{R}^3} d^3x \frac{4\sin^2 \xi}{(1+|u|^2)^2} \left( (1+\Omega)u_i-i\epsilon_{ijk}\xi_ju_k \right) \left( (1+\Omega)\bar{u}_i-i\epsilon_{ilm}\xi_l\bar{u}_m \right) \nonumber \\
&+&12\int_{\mathbb{R}^3} d^3x \frac{i\sin^2\xi}{(1+|u|^2)^2} \epsilon_{ijk}\xi_iu_j\bar{u}_k \geq 12\int_{\mathbb{R}^3} d^3x \frac{i\sin^2\xi}{(1+|u|^2)^2} \epsilon_{ijk}\xi_iu_j\bar{u}_k = 12 \pi^2 B.
\eea
The saturation occurs when the Bogomolny equations are obeyed 
\bea
(1+\Omega) \xi_i &=& 2i\frac{\sin^2 \xi}{(1+|u|^2)^2} \epsilon_{ijk}u_j\bar{u}_k, \\
 (1+\Omega)u_i &=& i\epsilon_{ijk}\xi_ju_k,
\eea
and the complex conjugated of the last formula. These are the Bogomolny equations previously derived in the language of the eigenvalues. Hence, we know that this set of equations (remember the background function is assumed to be (\ref{Omega-1})) has at least one solution. It describes the spherically symmetric situation when positions of the Skyrmion and the impurity coincide. Surprisingly, it is the unique solution. A probable reason for that may be a very restrictive nature of the nine Bogomolny equations.

First of all, derivatives of the fields obey the following relations 
\be
\xi_iu_i=\xi_i\bar{u}_i=0, \;\;\;\;\;\; u_i^2=\bar{u}_i^2=0.
\ee
Next, one finds that
\be
(1+\Omega)u_i\bar{u}_i= \pi^2 \frac{(1+|u|^2)^2}{\sin^2 \xi} \mathcal{B}_0, \;\;\;\; (1+\Omega)\xi_i^2=2\pi^2 \mathcal{B}_0,
\ee
and
\be
\xi_i^2=(1+\Omega)^2.
\ee
All that implies that the baryon charge density (and therefore the energy density) of any solution of the Bogomolny equation is completely determined by the form of the impurity
\be
\mathcal{B}_0=\frac{(1+\Omega)^3}{2\pi^2}.
\ee
As a result, there is no nontrivial moduli space. The self-dual Skyrmion cannot be taken away from the impurity. 
\subsection{Higher charge BPS Skyrmions}

It is possible to find background fields which support a higher charge BPS Skyrmion. An important family of Skyrme configurations is provided by the so-called rational map approximation \cite{Houghton1998}. Such Skyrme fields are defined by formulae (\ref{U})-(\ref{n}) with the profile function depending only on the radial coordinate, $r$, while the complex field $u(z)=R(z)$ is a rational map $R$ between Riemann spheres in terms of the complex coordinate $z = e^{i \phi} \tan \frac{\theta}{2}$. 

Although rational map Skyrmions do not solve the minimal Skyrme model for $|B|>1$, they have been widely studied providing a remarkably good approximation for the true solutions. They not only give quite accurate energy but also capture the geometry of the energy minima in each topological sector. Therefore, it seems of interest to see whether the considered here background field can transform a rational map Skyrmion into a self-dual (BPS) solution in the background field deformed model.  

In this framework, the eigenvalues $\lambda_i$ read
\be
\lambda_1 = -\xi_r(r), \qquad \qquad \lambda_2 = \lambda_3 = \frac{\sin \xi}{r} \frac{1+|z|^2}{1+|R|^2} \left| \frac{dR}{dz}\right|,
\ee
allowing a general study for any kind of rational map.

From these expressions and considering the BPS equations, it is manifest that for a well-behaved background field at $r = 0$, it is necessary that either the radial part of $\lambda_2$ (and $\lambda_3$) vanishes at the center or that the angular contribution is a constant. In fact, although the latter is the case of the charge 1 Skyrmion (given by the map $R=z$), it seems more convenient to consider a profile such that $\sin \xi/r \rightarrow 0$ when $r \rightarrow 0$ at the same time that we keep the well-studied rational maps. In this way, we ensure that the symmetries of the higher charge Skyrmions given by the rational map ansatze are not spoilt. In fact, in the original approach the profile function $\xi(r)$ is also found {\it a posteriori} by minimizing the energy functional once the rational map has been energetically determined (see \cite{Houghton1998} for details).

To achieve this desired behavior, at least one of the background fields needs to have an angular dependence, let us say $\Omega_1$. In particular, we can take
\be
1+\Omega_2 = 1+\Omega_3 \equiv 1+\Omega(r), \qquad \qquad 1+\Omega_1 \equiv (1+\Omega(r)) \, \Theta(\theta,\phi),
\ee
which by the BPS equations imply that
\be
\lambda_1^2= \xi_r^2 = (1+\Omega)^2, \qquad \qquad \lambda_2^2 = \lambda_3^2 = \frac{\sin^2 \xi}{r^2} \left(\frac{1+|z|^2}{1+|R|^2} \left| \frac{dR}{dz}\right| \right)^2 = (1+\Omega)^2 \, \Theta.
\ee

Hence, we can define the angular part of the background field as
\be \label{Theta}
\Theta(\theta,\phi) = \frac{1}{n^2} \left(\frac{1+|z|^2}{1+|R|^2} \left| \frac{dR}{dz}\right| \right)^2 ,
\ee
where $n$ is an arbitrary constant which seems natural to take it equal to the baryon charge of the BPS Skyrmion, {\it i.e.}, $n = B$. Therefore, for the radial contribution we are left just with 
\be
(1+\Omega(r))^2 = \xi_r^2(r) = \frac{B^2 \sin^2 \xi}{r^2},
\ee
which is satisfied by the profile function
\be
\xi (r) = 2 \arctan \left(\frac{r_0}{r} \right)^B,
\ee
with $r_0>0$ an arbitrary constant.

Then, it is easy to see that the radial background field for baryon charge $B$ is given by
\be
1+\Omega(r) = -\xi_r = \frac{2 B r_0^B}{r_0^{2 B}+r^{2 B}} \, r^{B-1},
\ee
so when $B = 1$ we recover the isotropic case studied above. Otherwise, the impurity $1+\Omega$ is also well-defined now as for $r=0$ it tends to zero.

Finally, it is worth mentioning that the charge of the BPS Skyrmion is given by the particular choice of the rational map, so profiles with an integer power $n$ of $r$ different from $B$ may be also allowed [being $n$ the same as in (\ref{Theta})], that is to say,
\be
\xi (r) = 2 \arctan \left(\frac{r_0}{r} \right)^n.
\ee
The only restriction would be the case $n=1$ for $B>1$ because, due to the non-zero derivative of the corresponding profile at $r=0$, the background field $\Omega_1(r,\theta,\phi)$ is not well-defined at the origin.

\subsection{Relation with other self-dual Skyrmions}
The self-dual background field modification of the original minimal Skyrme model (\ref{model}) is not the unique one leading to the Bogomolny equations (\ref{BOG-Sk-Imp}). In fact, there is some freedom in the definition of the static energy functional. Another possibility is the one introduced in \cite{LF} and further exploited in \cite{Luiz&Leandro}, where a deformation of the Skyrme term is also allowed. In fact, as previously stated, it presents solutions which, under some conditions, are analogous to the isotropic case reported above. Then, the resulting model is
\be
E_\Omega = E_{2, \Omega}   + E_{4, \Omega}
\ee
where now
\be
E_{2, \Omega} = \int_{\mathbb{R}^3} \left( (1+\Omega_1) \lambda^2_1+(1+\Omega_2) \lambda^2_2+(1+\Omega_2) \lambda^2_3 \right)d^3 x,
\ee
\be
E_{4, \Omega}=   \int_{\mathbb{R}^3} \left( \frac{\lambda^2_2\lambda^2_3 }{1+\Omega_1}  +\frac{\lambda^2_3\lambda^2_1}{1+\Omega_2}   +\frac{ \lambda^2_1\lambda^2_2}{1+\Omega_3} \right)d^3 x.
\ee
We remark that, contrary to the previous background field model, this set-up gives self-dual as well as anti-self-dual Bogomolny equations. Hence, solitons and antisolitons have exactly the same energy. 

Such an energy functional is also in the spirit of that considered in \cite{iterated}, with an application to the construction of the moduli space for the kink-antikink collision in the $\phi^4$ theory.

Importantly, the corresponding Bogomolny equations are identical to eq. (\ref{BOG-Sk-Imp}) with an additional arbitrariness of the $\pm$ sign. Therefore,  in the pertinent topological sector, again only one solution exists and all negative results concerning the moduli space hold.  

In general, any background field deformation of the minimal Skyrme model which leads to Bogomolny equations such that the resulting eigenvalues are uniquely defined by the background field (impurity) implies a triviality of the moduli space.
\section{Background field deformations with nontrivial moduli space}
Here we present Skyrme type models which, after a specific coupling with background fields, do give rise to infinitely many solitonic solutions in an arbitrary topological sector. However, unlike the minimal Skyrme model considered above, in both examples the initial model possesses infinitely many physically non-equivalent solitonic solutions in the self-dual sector. 
\subsection{A modified minimal Skyrme model}
The model which we want to focus on is given by the following expression
\be
E=E_{2, \mathcal{U}} + E_4,
\ee
where $E_2$ is a modified Dirichlet term
\be
E_2=\int_{\mathbb{R}^3} -\frac{1}{2} \mbox{ Tr } (R_i R_i) \; \mathcal{U}^2 d^3 x,
\ee
and $E_4$ is the Skyrme term. Here $\mathcal{U}$ is a function of the trace of the Skyrme matrix field $U$. In a sense, it plays the role of a field dependent coupling function (dielectric function). Then, rewriting the static energy in terms of the eigenvalues we find 
\bea
E&=&\int_{\mathbb{R}^3} \left( \mathcal{U}^2( \lambda^2_1+\lambda^2_2+\lambda^2_3) + \lambda^2_1\lambda^2_2+\lambda^2_2\lambda^2_3 + \lambda^2_3\lambda^2_1\right)d^3 x \nonumber \\
&=& \int_{\mathbb{R}^3} \left(  ( \mathcal{U}  \lambda_1 \pm  \lambda_2  \lambda_3)^2 + (  \mathcal{U} \lambda_2 \pm  \lambda_3  \lambda_1)^2 + (  \mathcal{U}\lambda_3 \pm  \lambda_1 \lambda_2)^2 \right) d^3 x \mp 6 \int_{\mathbb{R}^3}  \mathcal{U} \lambda_1 \lambda_2 \lambda_3 d^3 x  \nonumber \\
& \geq & 6 \left| \int_{\mathbb{R}^3}  \mathcal{U} \lambda_1 \lambda_2 \lambda_3 d^3 x \right| = 12 \pi^2 \left\langle \mathcal{U} \right\rangle |B|.
\eea
where $\left\langle \mathcal{U} \right\rangle$ is the average value of $\mathcal{U}$ over the whole $\mathbb{S}^3$ target space. The bound is saturated if and only if the following Bogomolny equations are fulfilled 
\be
\mathcal{U}  \lambda_1 \pm  \lambda_2  \lambda_3 =0, \;\; \mathcal{U}  \lambda_2  \pm  \lambda_1\lambda_3 =0, \;\; \mathcal{U}  \lambda_3 \pm  \lambda_1  \lambda_2 =0.
\ee
This implies that 
\be
\lambda_1^2=\lambda_2^2=\lambda_3^2= \mathcal{U}^2.
\ee
Following our previous analysis, this set of Bogomolny equations for $\mathcal{U} = \frac{1}{\beta} (1-\cos \xi)$ has a topologically nontrivial solution
\be
\xi = 2\arctan \frac{\beta}{r}, \;\;\; u=\tan \frac{\theta}{2} u^{i\phi}.
\ee
Since the model is Poincare invariant, we have infinitely many solutions generated by the 3-dimensional translations. Hence, a moduli space obviously exists. It is worth noting that a similar situation would also arise by considering a modified Skyrme term instead.

Now, we couple background fields in such a way that the nontriviallity of the self-dual sector remains preserved. For that, let us consider the following background field deformation
\be
E=E_{2, \mathcal{U}, \Omega} +E_4 +E_{3, \mathcal{U}, \Omega}
\ee
where the background field deformed terms read
\be
E_{2, \mathcal{U}, \Omega}=\int_{\mathbb{R}^3}  \mathcal{U}^2 \left( (1+\Omega_1)^2\lambda^2_1+(1+\Omega_2)^2\lambda^2_2+(1+\Omega_3)^2\lambda^2_3 \right) d^3x,
\ee
\be
E_{3, \mathcal{U}, \Omega}=-2\int_{\mathbb{R}^3}  \mathcal{U} (\Omega_1+\Omega_2+\Omega_3) \lambda_1\lambda_2\lambda_3  d^3x.
\ee
This background field modified model has the topological bound 
\be
E \geq 6 \int_{\mathbb{R}^3} \mathcal{U} \lambda_1\lambda_2\lambda_3  d^3x  = 12 \pi^2 \left\langle \mathcal{U} \right\rangle B,
\ee
where the equality is achieved for solutions of the corresponding, background field modified, Bogomolny equations
\be
\mathcal{U}  (1+\Omega_1) \lambda_1 -  \lambda_2  \lambda_3 =0, \;\; \mathcal{U}  (1+\Omega_2)\lambda_2  -  \lambda_1\lambda_3 =0, \;\; \mathcal{U} (1+\Omega_3) \lambda_3 -  \lambda_1  \lambda_2 =0,
\ee
which lead to simple expressions for the eigenvalues
\be
\frac{\lambda_1^2}{\mathcal{U}^2}= (1+\Omega_2) (1+\Omega_3), \;\; \frac{\lambda_2^2}{\mathcal{U}^2}= (1+\Omega_1) (1+\Omega_3), \;\;\frac{\lambda_3^2}{\mathcal{U}^2}= (1+\Omega_1) (1+\Omega_2).
 \ee
 An example of (infinitely many) solutions can be constructed assuming that the background field has only two components, $\Omega_1(r)$ and $\Omega_2=\Omega_3 \equiv \Omega(r)$, which are subjected to a constraint. Indeed, a unit charge BPS hedgehog Skyrmion, $\xi=\xi(r)$, $u=\tan \frac{\theta}{2} e^{i\phi}$, obeys 
 \be
\frac{1}{1+\Omega(r)} \frac{d\xi}{dr}=-(1-\cos \xi), \;\;\; \frac{1}{r} \sin \xi = (1-\cos \xi) \sqrt{(1+\Omega_1) (1+\Omega) }.
 \ee
 The first equation can be brought, by a suitable transformation of the radial coordinate, to the previous non-impurity form
 \be
 (1+\Omega(r)) dr = dr' \;\;\; \Rightarrow \;\;\; \xi_{r'}=-(1-\cos \xi).
 \ee
Then, assuming that the background field $\Omega_1$ is such that 
\be
r' = r \sqrt{ (1+\Omega_1) (1+\Omega) },
\ee
also the second equation is transformed into the previous form. To conclude, this coordinate transformation leads to the solution $\xi(r')= 2\arctan \frac{\beta}{r'}$. Furthermore, the Bogomolny equation in the $r'$ variable coincides  with the non-impurity version and therefore admits a translation symmetry.  
\subsection{The BPS Skyrme model with the self-dual background field}
For the sake of completeness we show that the BPS Skyrme model admits a background field extension which preserves the BPS nature of the original theory. Moreover, the background field deformed Bogomolny equation still supports infinitely many Skyrmions with arbitrary value of the baryon index. Similar computations were presented for the BPS baby Skyrme model in \cite{susy}. The pertinent energy functional reads
\be
E_{BPS, \; \Omega}= E_{6, \Omega} + \tilde{E}_{3, \Omega}+E_0,
\ee
where $E_{6, \Omega} $ is the deformed sextic term
\be
E_{6, \Omega} =\int_{\mathbb{R}^3} d^3x (1+\Omega)^2 \lambda_1^2\lambda_3^2\lambda_2^2,
\ee
$E_0$ is the non derivative part containing the potential $\mathcal{U}$ 
\be
E_0=\int_{\mathbb{R}^3} d^3x \mathcal{U},
\ee
and $\tilde{E}_{3, \Omega}$ is a version of the topological term in the presence of the impurity 
\be
\tilde{E}_{3, \Omega}= -2 \int_{\mathbb{R}^3} d^3x \Omega \sqrt{\mathcal{U}} \lambda_1\lambda_2\lambda_3.
\ee
Then, the topological bound can be easily found
\bea
E&=&\int_{\mathbb{R}^3} d^3x \left( (1+\Omega) \lambda_1\lambda_2\lambda_3 - \sqrt{\mathcal{U}}\right)^2 + 2\int_{\mathbb{R}^3} d^3x \sqrt{\mathcal{U}} \lambda_1\lambda_2\lambda_3 \nonumber \\
&\geq& 2\int_{\mathbb{R}^3} d^3x \sqrt{\mathcal{U}} \lambda_1\lambda_2\lambda_3 = 4\pi^2 \left\langle \sqrt{\mathcal{U}} \right\rangle B.
\eea
The bound is saturated if and only if the following Bogomolny equation is satisfied
\be
(1+\Omega) \lambda_1\lambda_2\lambda_3 - \sqrt{\mathcal{U}}=0,
\ee
which is just the background field modification of the usual BPS Skyrme Bogomolny equation. To prove that this equation still possesses infinitely many solutions we apply the standard parametrisation. Then, it can be rewritten as 
\be
(1+\Omega) \frac{2i \sin^2 \xi}{(1+|u|^2)^2} \epsilon_{ijk} \xi_i u_j \bar{u}_k - \sqrt{\mathcal{U}}=0 \label{BPS-mod}.
\ee
For simplicity we assume that $\Omega = \Omega (r)$. Then, we can introduce a new radial coordinate $r'$ defined via the relation
\be
r'^2dr'= \frac{r^2dr}{1+\Omega(r)},
\ee
which brings eq. (\ref{BPS-mod}) into the original BPS Skyrme Bogomolny equation in $(r',\theta, \phi)$ coordinates. This is a higher dimensional counterpart of transformation discussed in \cite{SD-imp-3} and \cite{iterated}. Then one can use the volume preserving diffeomorphisms in the new coordinates, which are known to be symmetries of the Bogomolny equation, to obtain background field  deformed BPS Skyrmions with an arbitrary shape. 
\section{Summary}

We have shown that a non-self-dual (non-BPS) topological soliton of a field theory can be transformed into a self-dual soliton by the inclusion of self-dual background fields. This requires a certain modification of the considered model (energy functional) by a coupling of suitable background fields with the original (dynamical) fields. For a particular choice of the background fields, the resulting theory supports a self-dual soliton of a given  ({\it only one}) value of the topological charge. Such a soliton saturates the pertinent topological bound and obeys the corresponding Bogomolny equations. Therefore, we can conclude that any soliton can be uplifted to its self-dual counterpart if a properly chosen background field is added. 

As we mentioned in the Introduction, a change of the geometry of he base space may also bring a solitonic theory into a BPS regime. It would be interesting to study a possible relation behind these two approaches in detail.

As a particular example we considered skyrmions in the minimal Skyrme model where background fields supporting BPS solitons with arbitrary charge have been found. Of course, a given set of the background fields can give rise to only one BPS Skyrmion with a fixed baryon charge. Solitons with another value of the topological charge do not obey the Bogomolny equation and in consequence do not saturate the topological bound. Obviously, this construction is not limited to the minimal Skyrme model and one can apply it to other non-BPS solitonic theories as the baby Skyrme model or the Abelian Higgs model with a non-critical coupling.
 Also, even if we have considered symmetric backgrounds for simplicity, the same will apply to more general cases. In particular, it might be of interest for composite solitons, such as domain wall Skyrmions \cite{Bjarke2014}. In the last years, these composite configurations have drawn much attention in the realm of condensed matter systems \cite{Muneto2012,Kasamatsu2013,Bjarke2018,Cheng2019}, being recently observed in chiral magnets \cite{Nagase2020}.

Although the resulting self-dual Skyrmion obeys a set of background field modified Bogomolny equations, there is no nontrivial moduli space. Physical observables (energy and topological densities) are uniquely determined by the form of the background fields. Therefore, the obtained exact solution is the only one and represents a Skyrmion located on top of the impurity. Equivalently, we can say that the original (non-self-dual) Skyrmion forms a bound state with the background field which ultimately saturates the topological energy bound. Hence, there is a static force between the impurity and soliton which prevents effortless separation. Thus, no non-trivial zero mode exists. Furthermore, there is no restoration of the (generalised) translational invariance in the self-dual sector. 

This fundamentally differs from the self-dual background field deformation of theories supporting self-dual single soliton solutions (as for example the $\phi^4$ model in (1+1) dimensions). In this case, despite of the inclusion of a background field which explicitly breaks the translational symmetry, the self-dual sector enjoys a generalised translation invariance which amounts to an appearance of zero modes and, in consequence, a moduli space where a distance between the soliton and impurity can be arbitrary changed. Apparently, uplifting a non-SD soliton to its SD counterpart by the use of a background field, unlike the screening of the static inter-soliton force, leads to very restrictive Bogomolny equations. 

This difference is clearly seen in the BPS Skyrme model which admits a self-dual background field extension leading to infinitely many BPS solutions, carrying arbitrary value of the baryon charge, also in the presence of a fix impurity. Hence, a nontrivial moduli space exists. In addition, there are (infinitely many) zero modes, which may be related to generalized volume preserving diffeomorphisms. Of course, here the starting theory supports self-dual solitons. 

Undoubtedly, it would be interesting to better understand conditions which imply the triviality (or non-triviality) of the moduli space for a given set of Bogomolny equations. 

All that can be summarized as follows.  If the original theory is a non-BPS one, then the background field can transform {\it only one} topological soliton into a self-dual solution, which is a  {\it non-separable} soliton-impurity bound state. Thus, no non-trivial moduli space exists. On the other hand, if we start with a BPS theory, then the self-dual background field deformation keeps the self-dual property unchanged in {\it any} topological sector. Now, soliton and impurity form a state with zero binding. This implies that the position of the soliton with respect to the impurity can be arbitrarly changed. Hence, a nontrivial moduli space does exist. 

\vspace*{0.2cm}

It is a widely known fact \cite{jose-susy1}-\cite{bjarke-nitta} that the non-triviality of the BPS sector can be related to the existence of supersymmetric extensions of the original bosonic theory.
It also applies to theories with impurities \cite{susy}. Therefore, we expect that the considered here deformation of the Skyrme model possesses an $\mathcal{N}=1$ SUSY version. 

The results of our work may lead to some difficulties if one would like to use the self-dual background field framework for understanding the dynamics, especially an annihilation, of non-self-dual topological solitons (non-self-dual asymptotic states). In the optimal situation we would like to have a self-dual soliton (obtained from a non-self-dual one by an application of background fields) which could be {\it freely} moved between infinities, at least along a fixed curve. Then, a generalization to a self-dual solution describing soliton-(anti)soliton scattering is straightforward, see \cite{SD-imp-3}. Unfortunately, in our construction the position of the self-dual soliton is {\it fixed}. A possible improvement might be achieved by taking advantage of the route developed for magnetic Skyrmions where an external (background) gauge field plays the main role, see \cite{BS} and \cite{Nitta}. Undoubtedly, to answer the question whether this is an artefact of the particular background field deformation considered in the paper or it is a general feature of any background field model requires further investigations.

Another possibility is to accept the non-BPS nature of the asymptotic states (solitons at infinities) and add a background field which will screen the inter-soliton static force, exactly as in the usual self-dual background field framework. 

\section*{Acknowledgments}
We would like to thank Andrzej Wereszczynski and Christoph Adam for useful discussions and comments. CN is supported by the INFN grant 19292/2017 {\it Integrable Models and Their Applications to Classical and Quantum Problems}. KO acknowlegdes support from the National Science Centre, Poland (Grant MINIATURA 3 No. 2019/03/X/ST2/01690 and Grant No. NCN 2019/35/B/ST2/00059).

\end{document}